\documentclass[fleqn,twoside]{article}
\usepackage{espcrc2}

\usepackage{graphicx}
\usepackage[figuresright]{rotating}

\newcommand{\AmS}{{\protect\the\textfont2
  A\kern-.1667em\lower.5ex\hbox{M}\kern-.125emS}}

\title{Spectral analysis of a large sample of BeppoSAX Seyfert spectra
  with Comptonization models: Preliminary results}

\author{P.O. Petrucci\address[laog]{Laboratoire d'Astrophysique de
    Grenoble, BP 43, 38041 Grenoble Cedex 9, France}, M.
  Dadina\address[iasf]{ Istituto TESRE, Via Gobetti 101 40129 Bologna,
    Italy}}

\begin{document}

\begin{abstract}
We present preliminary results of the spectral analysis of a large
sample of Seyfert galaxies observed by BeppoSAX. The only selection
criterium was a sufficiently large S/N ratio ($>$10) in the PDS band
(12-200 keV) to allow good detection up to the highest energy. The
resulting sample is composed of 28 objects (17 Seyfert 1, 11 Seyfert 2)
and 50 observations. Our main effort here is to adopt Comptonization
models to fit the different spectra on a truly broad band basis
(0.1-200 keV). We use two distinct disc-corona configurations, an
anisotropic slab and an isotropic spherical one. We discuss the
distributions of the physical parameters, like temperature and optical
depth of the corona and the reflection component, among this sample. We
also discussed the existence (or inexistence) of correlations between
these parameters.
\end{abstract}

\maketitle

\section{The data}
We select a large sample of Seyfert galaxies observed by BeppoSAX. The
only selection criterium was a sufficiently large S/N ($>$ 10) in the PDS
band to allow good detection and good spectral fitting up to the highest
($\sim$200 keV) energy. The names, as well as the Seyfert types and the
numbers of BeppoSAX observations, are reported in Table 1 for each object
of our sample.
 
The LECS and MECS event files were downloaded from the BeppoSAX archives.
The spectral counts were extracted from a circular region of 4' for the
MECS and 4' or 8' for the LECS depending in the source statistics in the
LECS band. We used the data of the three (or two, for observations done
after May 1997) MECS units merged together to increase the S/N.
 
\begin{table}[!h]
\caption{List of the Seyfert galaxies of our sample, with their types and the
  number of observations. It contains 28 objects (17
  Seyfert 1-1.5 and 11 Seyfert 1.9-2) and corresponds to 50 BeppoSAX
  observations}
\begin{tabular}{lcc}
\hline
{\bf Name} & {\bf Type} & {\bf Number of obs.}\\
\hline
 ESO141-G55  &     1    &     1\\
 NGC 4593    &   1      &     1\\
 FAIRALL 9   &     1    &     1\\
 NGC 3783    &   1      &     1\\
MKN110       &    1     &     2\\
MR2251-17.8  &    1     &    2\\
 MCG-6-30-15 &    1.2   &     1\\
IC 4329A     &  1.2     &      5\\
MKN 509      &   1.2    &     2\\
NGC 7469     &   1.2    &     2\\
MKN766       &   1.5    &     2\\
NGC 3516     & 1.5      &    2\\
 NGC 5548    &  1.5     &     5\\
 MKN841      &    1.5   &     2\\
 NGC7213     &    1.5   &     3\\
 NGC 562a    &    1.5   &     1\\
 MCG 8-11-11 &    1.5   &     1\\
 NGC 5506    &    1.9   &     3\\
 NGC2992     &     2    &     2\\
 NGC 2110    &     2    &     2\\
 MCG-5-23-16 &   2      &     1\\
 MKN 1210    &     2    &     1\\
NGC7582      &    2     &     1\\
NGC7314      &    2     &     1\\
ESO103-G35   &    2     &     2\\
NGC6300      &    2     &     1\\
NGC4507      &    2     &     1\\
NGC3281      &    2     &     1\\
\hline
\end{tabular}
\end{table}
The PDS data have been reduced using the mission specific software
XAS. Variable Raise Time correction has been applied so to increase the
S/N ratio.  Source counts time series have been inspected visually to
identify and eliminate spurious spikes that could affect the spectral
shape especially at the lower end of the PDS useful energy band. The +off
and -off background spectra have been checked to test their
self-consistency and to exclude significant contamination from unknown
sources.

\section{Models and method used}
\subsection{The models}
The main effort in this work, in comparison to comparable studies done in
the past on BeppoSAX data (\cite{del03}, \cite{mal03}, \cite{ris02}) is
that we adopt precise thermal comptonization models to fit the different
spectra.
 
We also used two corona-disc configurations: a slab geometry (code of
Haardt \cite{haa94}) and a spherical (i.e. Isotropic) geometry (code of
Poutanen \& Svensson \cite{pou96}). Both codes derived the
angle--dependent spectra of the disk--corona system using an iterative
scattering method. The slab code also treats properly the scattering
anisotropy inherent to the slab geometry .
 
The physical parameters of these models are: the temperature of the
corona, $kT_e$, the corona optical depth $\tau$, the soft photon
temperature $kT_{bb}$ and the reflection normalization $R$.
 
Most of the objects present Warm Absorber features and/or a Soft excess
in the low part (i.e. in the LECS energy band) of their spectra. Good and
realistic fits require to take these features into account. Detailed
analyses of these different components are however beyond the scope of
this paper, which is instead focused on the high energy continuum of the
sources. Then, as a first approximation, we have added simple components
(edges, gaussian, blackbody...) to the primary continuum to reproduce the
main features present in the soft part of the spectra.
 
The use of the complete BeppoSAX energy band, and especially the soft
part, is very important since, in the case of thermal comptonization
models, soft and hard X-ray part of the spectra are linked. For example,
the presence of a soft excess have some impact on the fit of the 2-10 keV
X-ray continuum spectral shape, which strongly depends on $kT_e$ and
$\tau$.  On the other hand the high energy cut-off observed in the hard
X-ray/soft $\gamma$-ray in a large part of our objects also strongly
depends on $kT_e$. Thus the way we fit the soft excess may influence
strongly our high energy fits. A consistent picture then necessitates the
use of
\begin{figure}[!h]
\includegraphics[width=\columnwidth]{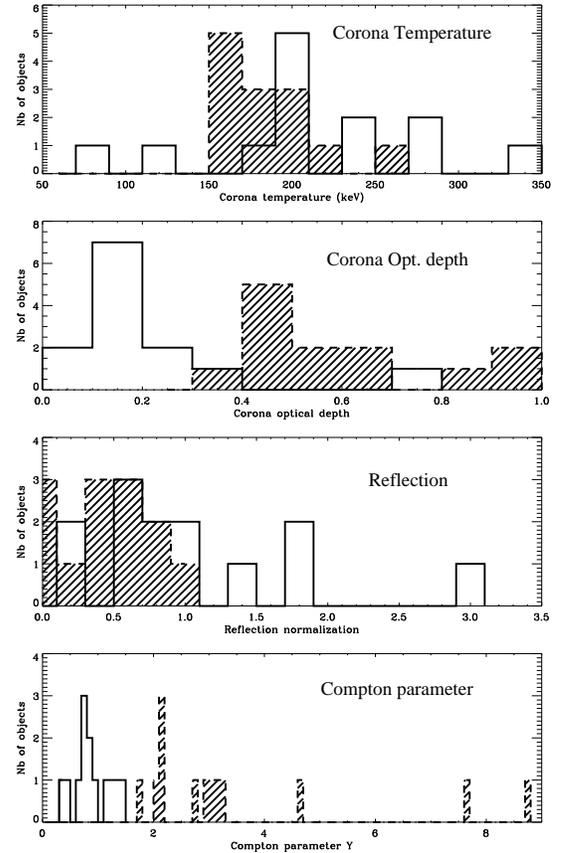}
\caption{Histograms of the corona temperature, corona optical depth,
  reflection normalization and compton parameter in the sub-sample
  already analyzed. The hatched histograms correspond to the sotropic
  geometry.}
\label{fig2}
\end{figure}
\begin{figure*}[!t]
\includegraphics[width=\textwidth]{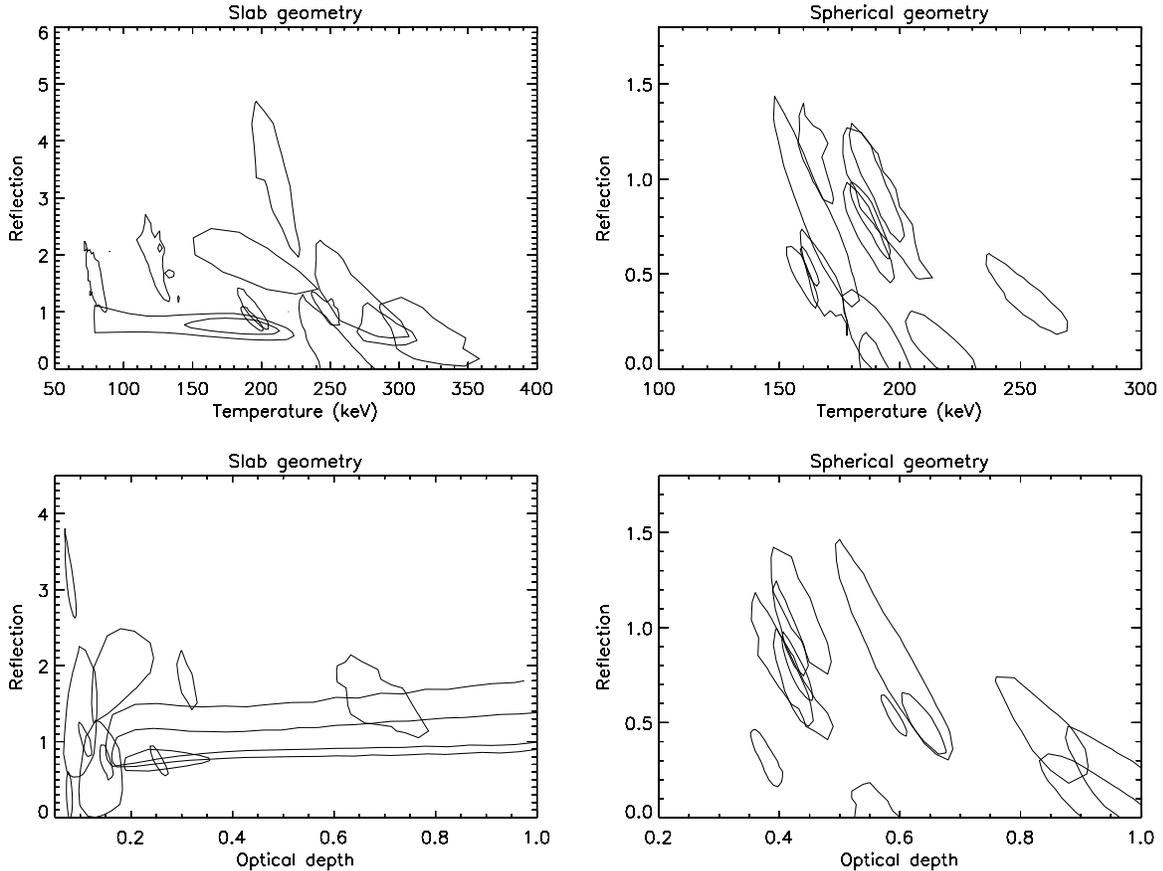}
\caption{Contour plots of the 13 observations of our subsample. We have
  plotted R vs kT$_e$ and R vs $\tau$ for the slab and isotropic
  geometries. The  contours correpond to the 90\% confidence level. }
\label{fig3}
\end{figure*}
the complete BeppoSAX energy interval.\\
This work can be seen as the continuation of the work already done for
some of the brightest sources of our sample (\cite{pet01}).

\subsection{The method}
For  each observation we check for spectral variability and treat
separatly the different spectral states 
 
Since we generally have very few constrains on the soft photon
temperature $kT_{bb}$, it is fixed in all fits to 10 eV. A first check
shows that the use of other values of $kT_{bb}$ does not change
significantly our results but more precise tests will be done in the
future.
 
All the parameters were let free to vary during the fit procedure but
most of them, $kT_e$, $\tau$ and $R$ apart, were fixed for the
computation of the contour plots. We use XSPEC (v11.2).

\section{Preliminary results}

Only 13 observations (6 objects) have been treated so far and thus the
preliminary results presented here have to be taken with caution:
\begin{itemize}
\item The two model geometry give acceptable fits with a total
  $\chi^2$/dof of 2096/1948 and 2156/1938 for the slab and spherical
  geometry respectively
\item Following the Kolmogorov-Smirnov test, the distribution of the best
  fit values of $\tau$ and $R$ significantly differ for the two geometry
  (cf. Fig. 1). Following the same test, the distributions of $kT_e$
  agree at more than 90\%. The resulting Compton parameter values are
  always larger for the spherical geometry. {\it The values of these
    different parameters thus appear to be strongly model dependent}
\item We do not find  clear correlation between the different parameters
   (using the Spearman (rank) and Pearson linear correlation test even if
   a anti-correlation between $R$ and $kT_e$ is suggested by the data
   (cf. Fig. 2) . The study of the complete sample will permit to
   conclude on more firm and solid statistics ground.  
\end{itemize}


\begin{thebibliography}{9}
\bibitem{del03}Deluit \& Courvoisier 2003 A\&A 399 77
\bibitem{haa94}Haardt 1994, PhD Thesis SISSA
\bibitem{mal03}Malizia et al. ApJL in press (astro-ph/0304133)
\bibitem{ris02}Risaliti 2002 A\&A 386, 379
\bibitem{pet01}Petrucci et al. 2001, ApJ, 556, 716
\bibitem{pou96}Poutanen \& Svensson 1996, ApJ, 470, 249
\end{thebibliography}
\end{document}